\documentclass[preprint]{aastex}

\newcommand{\lya}{Ly$\alpha$}

\begin{document}

\title{DEEP LBT/LUCI SPECTROSCOPY OF A LYMAN-$\alpha$ EMITTER CANDIDATE AT 
$z\simeq7.7$\footnotetext{
The LBT is an international collaboration among institutions in the United 
States, Italy and Germany. LBT Corporation partners are: The University of 
Arizona on behalf of the Arizona university system; Istituto Nazionale di 
Astrofisica, Italy; LBT Beteiligungsgesellschaft, Germany, representing the 
Max-Planck Society, the Astrophysical Institute Potsdam, and Heidelberg 
University; The Ohio State University, and The Research Corporation, 
on behalf of The University of Notre Dame, University of Minnesota and 
University of Virginia.}}

\author{Linhua Jiang\altaffilmark{1,2}, Fuyan Bian\altaffilmark{3},
Xiaohui Fan\altaffilmark{3}, Hannah B. Krug\altaffilmark{4},
Ian D. McGreer\altaffilmark{3}, Daniel P. Stark\altaffilmark{2,3},
Benjamin Cl{\'e}ment\altaffilmark{3}, and Eiichi Egami\altaffilmark{3}}
\altaffiltext{1}{School of Earth and Space Exploration, Arizona State
   University, Tempe, AZ 85287-1504, USA; linhua.jiang@asu.edu}
\altaffiltext{2}{Hubble Fellow}
\altaffiltext{3}{Steward Observatory, University of Arizona,
   933 North Cherry Avenue, Tucson, AZ 85721, USA}
\altaffiltext{4}{Department of Astronomy, University of Maryland, 
	College Park, MD 20742, USA}

\begin{abstract}

We present deep spectroscopic observations of a \lya\ emitter (LAE) candidate 
at $z\simeq7.7$ using the infrared spectrograph LUCI on the $2\times8.4$m 
Large Binocular Telescope (LBT). The candidate is the brightest among the 
four $z\simeq7.7$ LAE candidates found in a narrow-band imaging survey by 
\citet{krug12}. Our spectroscopic data include a total of 7.5 hours of 
integration with LBT/LUCI and are deep enough to significantly 
($3.2\sigma-4.9\sigma$) detect the \lya\ emission line of this candidate, 
based on its \lya\ flux $\rm 1.2\times10^{-17}\ erg\ s^{-1}\ cm^{-2}$ 
estimated from the narrow-band photometry. However, we do not find any 
convincing signal at the expected position of its \lya\ emission line, 
suggesting that this source is not an LAE at $z\simeq7.7$. 
The non-detection in this work, together with the previous studies of 
$z\simeq7.7$ LAEs, puts a strong constraint on the bright-end \lya\ luminosity 
function (LF) at $z\simeq7.7$. We find a rapid evolution of the \lya\ LF from 
$z\simeq6.5$ to 7.7: the upper limit of the $z\simeq7.7$ LF is more than 5 
times lower than the $z\simeq6.5$ LF at the bright end 
($f \rm \ge 1.0 \times 10^{-17}\ erg\ s^{-1}\ cm^{-2}$, or 
$L \rm \ge 6.9 \times 10^{42}\ erg\ s^{-1}$). This is likely caused by an 
increasing neutral fraction in the IGM that substantially attenuates \lya\ 
emission at $z\simeq7.7$.

\end{abstract}


\section{INTRODUCTION}

During the epoch of cosmic reionization, the intergalactic medium (IGM) was 
ionized by the first astrophysical objects, and the Universe became 
transparent to UV photons. Measurements of CMB polarization \citep{kom11} and
studies of high-redshift quasars \citep{fan06} have shown that reionization 
began earlier than $z\sim10$ and ended by $z\sim6$. The process of 
reionization is complex, and it is still not clear how and when it exactly 
occurred. Early observational studies were mostly based on high-redshift 
luminous quasars \citep[e.g.,][]{mes07,car10,mcg11}. In recent years, as the 
number of galaxies found at $z\ge6$ has increased dramatically, high-redshift 
galaxies have started to play an important role in the study of reionization.

In the last decade, a large number of \lya\ emitters (LAEs) and Lyman-break 
galaxies (LBGs) at $z\ge6$ have been discovered by two complementary methods:
the narrow-band technique and the dropout technique, respectively. 
In particular, the narrow-band (or \lya) technique has a high success rate of 
spectroscopic confirmation. Three atmospheric windows with little emission 
from OH sky lines in the optical are often used to search for galaxies at 
$z\simeq5.7$, 6.5, and 7. So far more than 200 LAEs have been 
spectroscopically confirmed at these redshifts
\citep[e.g.,][]{tan05,iye06,shi06,hu10,ouc10,kas11,rho12}.
Furthermore, a $z\simeq7.22$ LAE was confirmed recently \citep{shibuya12}, 
demonstrating the ability to detect distant galaxies with powerful 
ground-based telescopes.

Individual galaxies are usually too faint to provide useful information about 
the IGM ionization state during the reionization era. However, much 
can be learned from their statistical properties, such as the evolution of the 
\lya\ luminosity function (LF), and the fraction of galaxies with strong
\lya\ emission lines among LBGs \citep[e.g.,][]{kas11,sta11,treu12}. 
For example, recent studies have shown that the \lya\ LF evolves rapidly from 
$z\sim5.7$ to 6.5 \citep{hu10,ouc10,kas11}, while the evolution of the LAE UV 
LF is far less severe. This was 
explained by the increasing neutral fraction of the IGM from $z\sim5.7$ to 6.5 
that attenuates \lya\ emission via resonant scattering of \lya\ photons. 
As such, one would expect a further decline of the \lya\ LF at $z>6.5$. 
Although the $z\sim7$ LF has not been well determined due to the limited 
number of LAEs known at $z\sim7$, the current data do seem to indicate such an 
evolutionary trend \citep{ota12,shibuya12}. 

The narrow-band technique is now being used to search for higher redshift LAEs 
at $z\simeq7.7$ \citep[e.g.,][]{hib10,til10,cle12,krug12}. This redshift 
corresponds to an OH-dark window at $\sim$1.06 $\mu$m in the near-IR. 
Each of these narrow-band surveys, except the one by \citet{cle12}, found 
several candidates. Surprisingly, they also found that the \lya\ LFs at 
$z\simeq7.7$ derived from their photometric samples are consistent with the LF
at $z\simeq6.5$. On the other hand, none of their LAE candidates have been 
spectroscopically confirmed. Spectroscopic identification of these objects is 
very challenging, but is critical for understanding the properties of LAEs and 
the IGM state during reionization. In this paper we present 
deep LBT/LUCI spectroscopy of a $z\simeq7.7$ LAE candidate found by
\citet{krug12}. In Section 2 we describe our 
observations and data reduction. We then present our results in Section 3, and 
discuss the results in Section 4. We use a $\Lambda$-dominated flat cosmology 
with $H_0=70$ km s$^{-1}$ Mpc$^{-1}$, $\Omega_{m}=0.3$, and 
$\Omega_{\Lambda}=0.7$.

\section{OBSERVATIONS AND DATA REDUCTION}

Our target (hereafter LAEz7p7) is a LAE candidate at $z\simeq7.7$ selected 
from \citet{krug12}, who carried out a deep imaging survey of 
$z\simeq7.7$ LAEs with two ultra narrow-band filters centered at 1.056 and
1.063 $\mu$m. The filter width is $\sim8$ \AA. The survey volume is 
$2.8\times10^4$ Mpc$^3$, and the depth is 22.4 AB mag (50\% completeness)
in the narrow band, corresponding to a \lya\ flux limit of
$\rm 0.8\times10^{-17}\ erg\ s^{-1}\ cm^{-2}$. They found four candidates down 
to the survey limit in the two bands. LAEz7p7 is the brightest one (21.87 AB 
mag in the $1.056\mu$m band) with expected \lya\ flux of 
$\rm (1.21\pm0.16) \times 10^{-17}\ erg\ s^{-1}\ cm^{-2}$.

We observed LAEz7p7 using the infrared spectrograph LUCI \citep{age10}
on the $2\times8.4$m LBT on 11 December 2012 and 4 March 2013 (UT). The 
observing conditions were good, with mostly clear sky and decent seeing 
($0\farcs6\sim0\farcs9$). The observations were made in longslit mode with a 
$1\arcsec$ slit. We did a blind offset from a bright star to LAEz7p7. 
The long slit had a position angle of 53.08 deg so that it covered another 
nearby star with $J_{\rm Vega}=18.5$ mag. This star (hereafter RefStar)
is used as a reference object 
(and as a standard star for flux calibration) during the data reduction. 
The left panel of Figure 1 shows the positions of these 
objects in the LAEz7p7 field. We chose to use the second order of the 200 H+K 
grating with a wavelength coverage of $\sim4000$ \AA,
centered at 1.1 $\mu$m. This provides a resolving power $R\sim1000$.
The exposure time for each science image was 900 seconds. The individual 
exposures were dithered along the slit. The total integration
time was 7.5 hours. We also observed a standard star UKIRT FS128
with the same configuration for the purpose of flux calibration.

We reduced the LUCI data using standard methods based on our own customized 
pipeline \citep{bian10}. The basic procedure is as follows. For each raw 
science image, we first generated a 2-D wavelength map using many OH sky 
lines in the image. 
The closest dithered 
neighboring frame was then subtracted from this image. This step removed the
dark current and some instrument-related artifacts, and also performed the 
first sky subtraction. Next, the image was flat-fielded by a master flat image 
that was made from a series of flat images taken in the same night. A more 
accurate sky subtraction was performed with the \citet{kel03} algorithm.
Finally we rectified the reduced 2-D images (raw images were curved), and 
interpolated them onto a uniform rectangular output wavelength grid based on 
the individual 2-D wavelength maps.

Before we stacked these images, we shifted them 
along the spatial direction so that the bright RefStar has 
the same position in all the images. The accurate position of RefStar
in each image was determined by fitting a Gaussian profile to 
the average flux profile along the spatial direction. The images were then 
weighted on a frame-by-frame basis by sky transparency, seeing, and background 
variance. The transparency was measured by the relative flux of RefStar.
The variation of the transparency during the same run was very small 
(a few percent). Seeing was measured as the FWHM of the PSF from RefStar. 
The variance of background was calculated around the 
expected position of LAEz7p7. A final combined 2-D image was created 
by stacking the weighted science images with a sigma ($5\sigma$) rejection.
The PSF FWHM of RefStar in the stacked image is $0\farcs7$, well
consistent with those in input images.

\section{RESULTS}

Our main result is that we do not detect any convincing signal at the expected 
position of the LAEz7p7 \lya\ emission line in our stacked spectroscopic 
image. As we will see, this image is deep enough to detect LAEz7p7 at a 
significance level of $3.2\sigma-4.9\sigma$, if it is a $z\sim7.7$ LAE with 
the expected \lya\ flux given by \citet{krug12}.
In Figure 1, the right panel shows part of the stacked spectroscopic image.
With the known pixel scale ($0\farcs249$) 
and wavelength map, the expected position of LAEz7p7 in the 2-D image
is accurately determined. This position is in the middle of the 
box in the right panel. The box covers a range of 50 \AA\ centered at 1.056 
$\mu$m. A zoomed-in version is shown in Figure 3(e), and the extracted 1-D 
spectrum is given in Figure 2. Close visual inspection shows that there is no 
sign of any significant detection at this position.

We determine the absolute flux calibration using the standard star and 
RefStar. Their 1-D spectra are extracted within a 7-pixel ($1\farcs75$) window 
using IRAF task {\tt apall}. The standard star UKIRT FS128 is a M5V star with 
$J_{\rm Vega}=13.00$ mag. The system response curve is derived by comparing 
the observed spectrum of FS128 with the model spectrum of a M5V star. The 
spectrum of RefStar is corrected with the response curve, and scaled to match 
its $J$-band photometry. The flux calibration (or conversion between counts 
and flux) derived from the two stars is consistent with each other ($<0.15$ 
mag), indicating that RefStar (and thus LAEz7p7) was placed in the middle of 
the slit. We calculate the background variance around the expected position of
LAEz7p7, and convert it to a $1\sigma$ detection per pixel using 
the absolute flux calibration. This $1\sigma$ limit is used to estimate the 
detection limit of our image. In order to do so, we generate a 2-D model \lya\ 
spectrum for LAEz7p7, based on the 1-D composite spectrum of $z\simeq6.5$ LAEs 
given by \citet{kas11}. The model spectrum has the same pixel scale 
as our image. Its PSF FWHM is also $0\farcs7$. When extracted to 1-D spectrum, 
the model spectrum has the shape of the composite $z\simeq6.5$ spectrum.
The model spectrum is shown in Figure 3(a).

The detection significance of a line depends on the number of pixels extracted.
We calculate it in a box of 8 \AA\ (the filter width) by 
$1\arcsec$ (4 pixels). Based on the $1\sigma$ limit derived above, we find 
that a line with $\rm 1.21\times10^{-17}\ erg\ s^{-1}\ cm^{-2}$ should be 
detected at a $4.9\sigma$ level at its expected position in our image.
Another way to estimate the detection limit is commonly used for detecting 
faint objects in 2-D imaging data. We assume that the model LAE is detected 
at $\ge1.5\sigma$ in each of 5 contiguous pixels (the 5 brightest 
pixels here). In this case we expect a $3.2\sigma$ detection for LAEz7p7. 
This limit ($3.2\sigma$) is lower than $4.9\sigma$ derived above. This is 
due to the low resolution and good seeing, which cause the flux to be highly 
concentrated in the central pixels.

Visual inspection is another powerful way to identify faint objects.
We scale the model LAE and place it at the expected position of LAEz7p7. 
We then check whether we are able to identify it. The results 
are shown in Figure 3. Panel (e) shows the expected position of LAEz7p7, i.e., 
the central part of the box in Figure 1. The model \lya\ line is scaled to be 
5, 2, and 1 times the estimated \lya\ flux of LAEz7p7. The scaled 
spectra are placed at the expected position, shown in 
panels (b), (c), and (d), respectively. The images on the right-hand side 
are the Gaussian smoothed images. 
The figure demonstrates that LAEz7p7 can be clearly identified, if its \lya\ 
emission line is as strong as the value given by \citet{krug12}.
In fact, if LAEz7p7 is a real LAE, its \lya\ flux is likely higher than this
value, because the 1.056 $\mu$m filter is very narrow, and is very 
likely to only cover part of the \lya\ line.

We emphasize that the LAEz7p7 nominal position during our observations was 
securely located within the slit, with minimum slit losses owing to excellent 
seeing. This is ensured by two facts. One is the consistency of the flux 
calibration between RefStar and the standard star, as we discussed 
earlier. The other fact is the detection of a very faint galaxy
($J_{\rm Vega}=22.6$ mag) in the same slit. This object was well aligned with 
LAEz7p7 and RefStar, and was serendipitously covered by the slit
(faint galaxy in Figures 1 and 2). We detect both line and continuum 
emission for the galaxy. This ensures that LAEz7p7 was in the middle of the 
slit. In addition, the detection of such a faint object (Figure 2) indicates 
that the image depth is very good.

\section{DISCUSSION}

With our deep LBT/LUCI spectroscopy, we have ruled out the existence of a 
strong emission line at $\lambda\sim1.056$ $\mu$m in LAEz7p7, although this is
not sufficient for us to fully explain the nature of this object.
The photometric samples of $z\simeq7.7$ LAEs (and all other high-redshift 
galaxies), especially those with single-band detections, could be contaminated 
by various interlopers, such as low-redshift red or dusty galaxies, Galactic 
late-type dwarf stars, variable (or transient) objects, and even noise spikes.
The contamination in LAE samples has been discussed in the previously 
mentioned LAE papers \citep[also see e.g.,][]{pir13}.
Because of the rarity of $z\simeq7.7$ galaxies, even if the vast majority of 
contaminants can be removed with careful selection, a tiny fraction of them
could dominate a sample, highlighting the importance of spectroscopic 
follow-up observations.

The non-detection of LAEz7p7 allows us to put a constraint on the bright-end
\lya\ LF at $z\simeq7.7$. The \citet{krug12} survey has a volume of 
$2.8\times10^4$ Mpc$^3$, and found one candidate LAE (LAEz7p7) with \lya\ flux 
higher than $\rm 10^{-17}\ erg\ s^{-1}\ cm^{-2}$. This translates to an upper 
limit of the spatial density at the bright end
($L \rm \ge 6.9 \times 10^{42}\ erg\ s^{-1}$). Figure 4 shows the upper limit
(red circle), compared to previous results. The dashed blue and green lines
illustrate the strong decline of the bright-end \lya\ LF from
$z\simeq5.7$ to 6.5 \citep{kas11}. This has been explained by 
an increasing neutral fraction of the IGM. As such, we 
expect to see further decline of the \lya\ LF towards higher redshift. 
The constraints from the $z=6.96$ LAE \citep[magenta circle;][]{iye06}, from
the $z=7.22$ LAE \citep[cyan circle;][]{shibuya12}, and from LAEz7p7 suggest 
such a trend between $z\simeq7$ and 7.7. Note that the above LAEs at $z=6.96$
and 7.22 were discovered in much larger survey volumes.

The combination of this work and previous work enables us to put a more
stringent constraint on the \lya\ LF. \citet{cle12} did not 
find any $z\simeq7.7$ LAEs in their survey. Although \citet{hib10} found 7 
LAE candidates, none of their 5 brightest candidates was confirmed by 
follow-up spectroscopy, as mentioned by \citet{cle12}. A conservative summary 
is that, in the above three surveys, including \citet{krug12}, there is no 
$z\simeq7.7$ LAE with \lya\ emission higher than 
$\rm 10^{-17}\ erg\ s^{-1}\ cm^{-2}$
($L \rm \ge 6.9 \times 10^{42}\ erg\ s^{-1}$). 
If we assume a non-evolution of \lya\ LF from $z\simeq6.5$ to 7.7, we expect 
to find $\sim6$ LAEs down to the above flux limit, based on the LF of 
\citet{kas11}. The probability of finding zero LAEs is 0.0025, under the 
assumption of a Poisson distribution. By combining the three surveys, the
uncertainty due to cosmic variance has been largely reduced, and the Poisson
statistical uncertainty dominates the uncertainty of the LF.
Note that there is a discrepancy between the \citet{kas11} LFs and the
\citet{hu10} LFs, but they are roughly consistent at the brightest end.
The non-detection in the three studies translates to an upper limit of 
the bright-end \lya\ LF at $z\simeq7.7$. The black circle in Figure 4 
represents this upper limit (corresponding to the detection of one LAE). 
This limit is much stronger than that from LAEz7p7 alone. It shows a rapid 
evolution of the \lya\ LF at the bright end from $z\simeq6.5$ to 7.7: the 
upper limit of the $z\simeq7.7$ LF is a factor of 6 times lower than the 
$z\simeq6.5$ LF.

The decline of the \lya\ LF does not necessarily mean an increasing fraction
of neutral IGM, since this could also be caused by the intrinsic luminosiy 
and/or density evolution of LAEs \citep[e.g.,][]{jen13}. With limited 
information, we cannot rule out any intrinsic evolution. 
In particular, we know that the UV LF of LBGs evolves at this redshift 
\citep[e.g.][]{bou12,oes12}. However, such a rapid change of the LF in Figure 
4 is not likely to be due mainly to intrinsic evolution, because the time 
interval between $z=6.5$ and 7.7 is only 165 Myr, and the physical properties 
of LAEs evolve slowly at $3\le z\le6$ \citep[e.g.,][]{mal12,jiang13}. 
On the other hand, the increasing optical depth of the \lya\ forest towards
$z>6$ quasars indicates a rapid rise in the IGM neutral fraction. 
Because of radiative transfer effects, the \lya\ emission line is
largely shaped by its environment and more attenuated by a higher fraction of
neutral hydrogen \citep[e.g.,][]{mal04,zheng10,dij11}. 
At $z\simeq6.5$, the neutral fraction is still very small 
($\ll 10$\%) \citep{fan06}, yet it may already cause strong suppression on the 
\lya\ emission (Figure 4). A recently discovered quasar at $z=7.08$ 
\citep{mor11} was estimated to have a neutral fraction of $\sim$10\% in the
surrounding IGM \citep{bol11}. 
We thus expect a much higher neutral fraction of the IGM at 
$z\simeq7.7$ than that at $z\simeq6.5$. Therefore, the \lya\ emission of 
$z\simeq7.7$ LAEs would be strongly suppressed, which is broadly consistent 
with the observational trend in Figure 4.

Finally, we estimate the fraction of observed \lya\ flux that is reduced by 
neutral IGM at $z\simeq7.7$. We assume a pure luminosity evolution of the 
\lya\ LF from $z\simeq6.5$ to 7.7. We also assume that such evolution was
caused by the IGM. The black line in Figure 4 shows the evolved $z\simeq6.5$ 
LF matched to the LF upper limit (black circle) at $z\simeq7.7$.
We find that the \lya\ luminosity at $z\simeq7.7$ is reduced by 
at least a factor of two compared to that at $z\simeq6.5$.

\acknowledgments

Support for this work was provided by NASA through Hubble Fellowship grant
HST-HF-51291.01 awarded by STScI,
which is operated by the Association of Universities for Research in
Astronomy, Inc., for NASA, under contract NAS 5-26555. FB, XF, and IDM 
acknowledge supports from NSF grants AST 08-06861 and 11-07682.

{\it Facility:}
\facility{LBT (LUCI)}

\clearpage
\begin{figure}
\epsscale{1}
\plotone{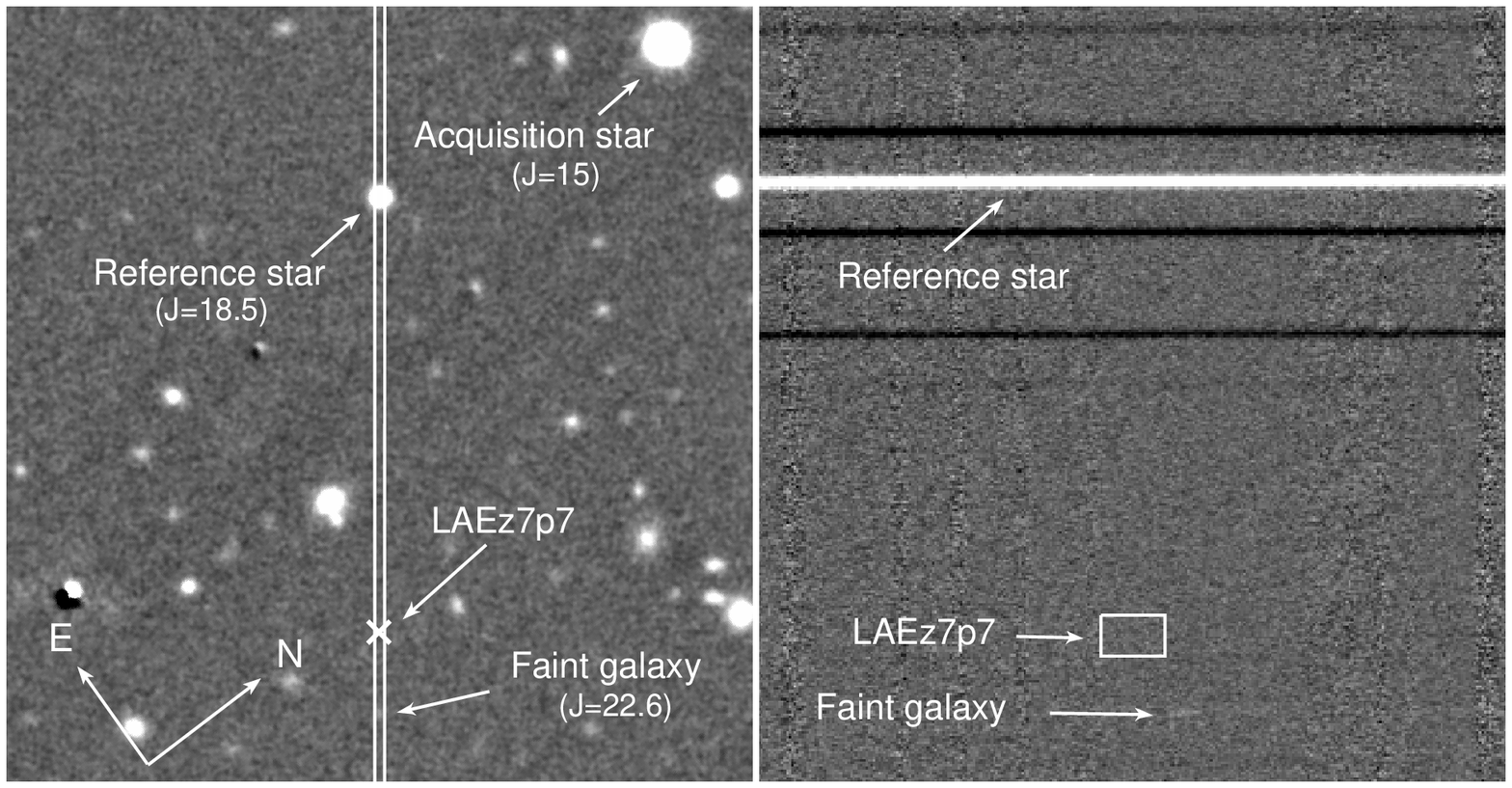}
\caption{Left: Deep $J$-band image of the LAEz7p7 field from \citet{krug12}. 
The position of LAEz7p7 is marked as the cross. With a position angle of 
53.08 deg, the long slit simultaneously covered LAEz7p7, the reference star 
RefStar ($J_{\rm Vega}=18.5$), and a very faint galaxy ($J_{\rm Vega}=22.6$ 
mag). Right: Part of the combined 2-D spectrum ($1.026-1.083$ $\mu$m) 
from LBT/LUCI.
We clearly detect the line and continuum emission for the faint galaxy. The 
box is centered at 1.056 $\mu$m, with a size of 50 \AA\ by $4\arcsec$. The 
\lya\ emission line of LAEz7p7 is expected to be in the middle of the box. 
However, we do not see any sign of detection. The central part of the box is 
zoomed in in Figure 3(e).}
\end{figure}

\begin{figure}
\epsscale{0.6}
\plotone{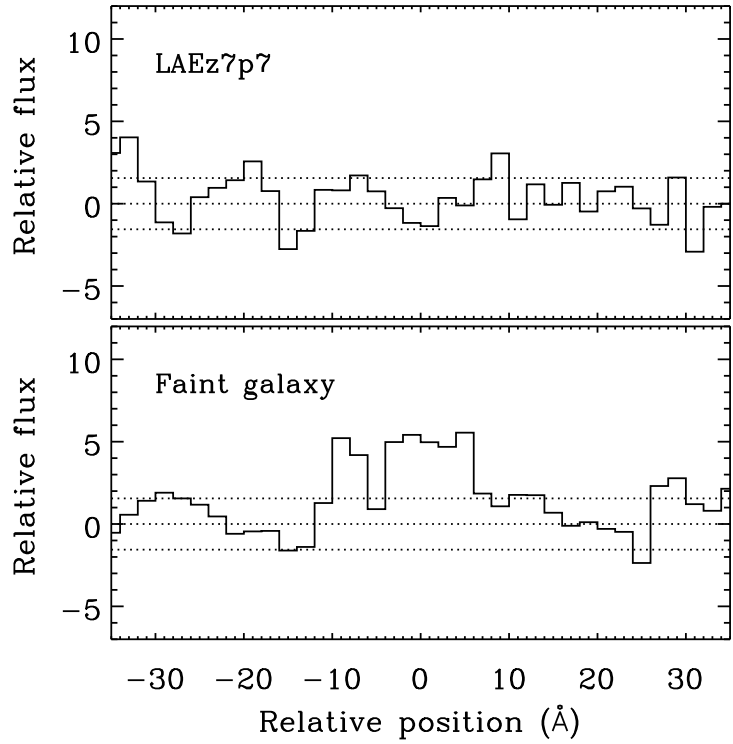}
\caption{1-D spectra of LAEz7p7 and the faint galaxy. The spectra are 
extracted within a 5-pixel ($1\farcs25$) window using IRAF task {\tt apall}.
The dotted lines represent zero flux and $1\sigma$ deviation. The spectrum of
LAEz7p7 is centered at 1.056 $\mu$m, and does not show any significant signal.
The spectrum of the faint galaxy is centered at 1.060 $\mu$m (as indicated in
the right panel of Figure 1), and shows a strong detection.}
\end{figure}

\begin{figure}
\epsscale{0.6}
\plotone{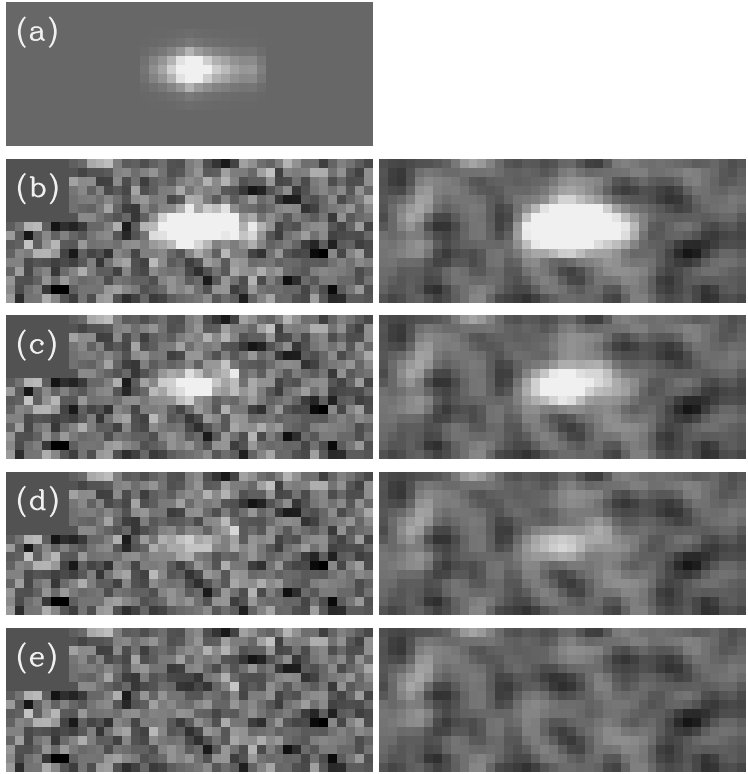}
\caption{\lya\ emission lines of a model LAE at the expected position of 
LAEz7p7 in our 2-D spectroscopic image. Panel (a) shows the model LAE. 
Panel (e) shows the expected position of LAEz7p7 (the central part of the box 
in Figure 1). The model line is scaled to be 5, 2, and 1 times the estimated 
\lya\ flux of LAEz7p7. The scaled 
lines are then placed at the expected position. The results are 
shown in (b), (c), and (d). The images on the right-hand side are the 
Gaussian smoothed version (with $\sigma=1$ pixel)
of the images on the left-hand side. 
It is clear that we are able to visually identify LAEz7p7, if its \lya\ 
emission is as strong as expected.}
\end{figure}

\begin{figure}
\epsscale{0.8}
\plotone{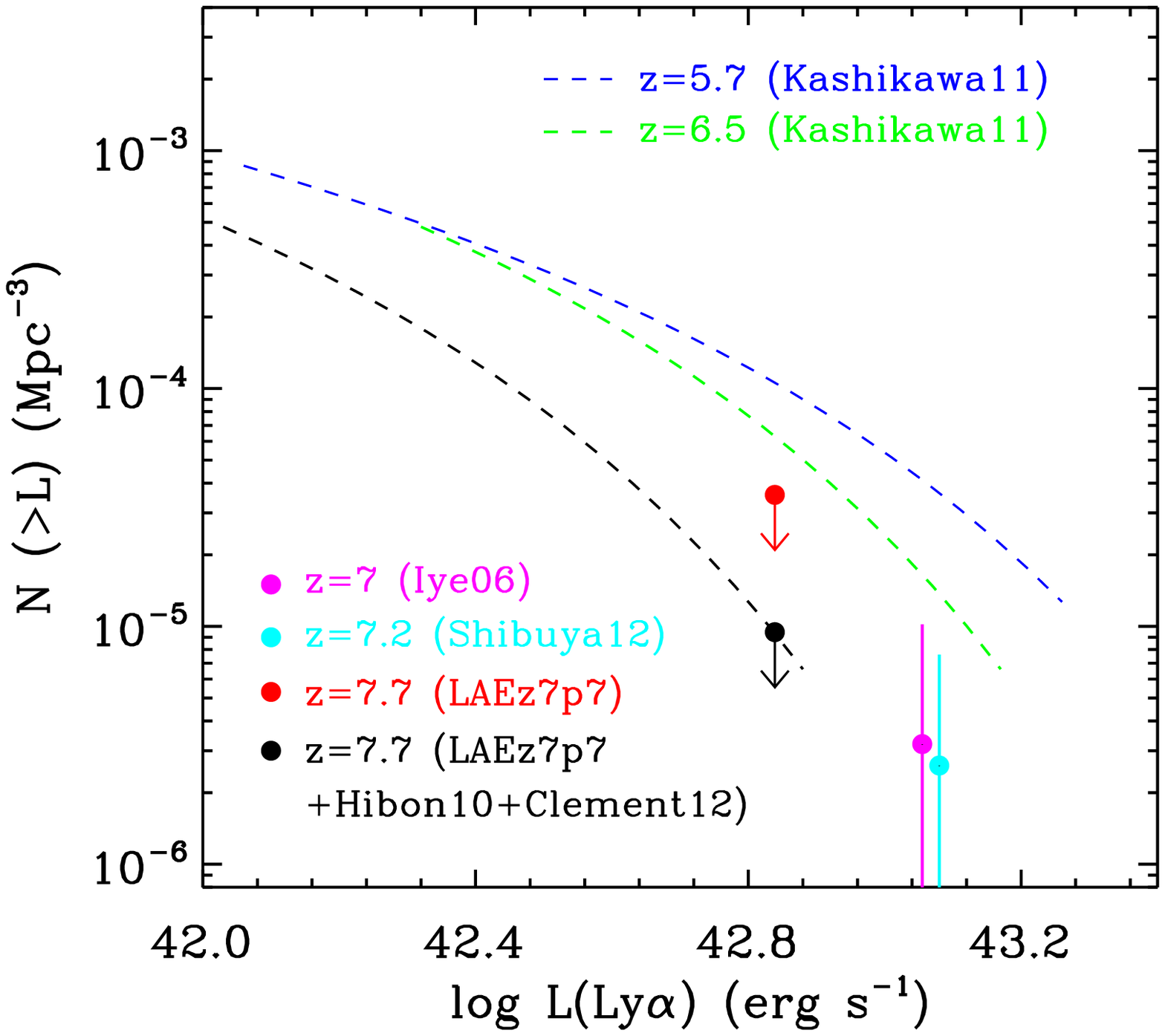}
\caption{Observed \lya\ LFs at high redshift. The dashed blue and green lines 
represent the LFs at $z\simeq5.7$ and 6.5 from \citet{kas11}. The luminosity 
coverage of the two lines reflects the actual luminosity coverage in their
sample. The magenta and cyan circles 
represent the LFs at $z\simeq7$ and 7.2, respectively \citep{iye06,shibuya12}.
The red and black circles indicate the upper limits of the bright-end LF 
at $z\simeq7.7$, derived from the non-detection of LAEz7p7 and from the 
combination of this work and previous studies \citep{hib10,cle12}. 
The black dashed line shows the evolved $z\simeq6.5$ LF matched to the upper 
limit of the $z\simeq7.7$ LF (black circle). The figure shows a strong
evolution of the \lya\ LF at the bright end from $z\simeq6.5$ to 7.7.}
\end{figure}

\end{document}